\documentclass[%
 reprint,
superscriptaddress,
 amsmath,amssymb,
 aps,
]{revtex4-2}

\usepackage{graphicx}
\usepackage{dcolumn}
\usepackage{bm}

\usepackage{physics}
\usepackage{hyperref}
\usepackage{cleveref}
\usepackage{mathrsfs} 
\usepackage{booktabs}
\usepackage{multirow}
\usepackage{makecell}
\usepackage[caption=false]{subfig} 
\usepackage{longtable}
\usepackage{upgreek}
\usepackage{xcolor}
\usepackage{float}
\usepackage{verbatim}

\newcommand{\figref}[1]{Fig.~\ref{#1}}

\begin{document}
\preprint{ADP-23-27/T1236}

\title{How to understand the $\rho$ resonance from the quark model and $\pi\pi$ $P$-wave phase shift}

\author{Wen-Ze Zhao}%
\affiliation{%
School of Physical Sciences, University of Chinese Academy of Sciences, Beijing 100049, China
}%

\author{Ru-Hui Ni}%
\email{corresponding author: niruhui@ucas.ac.cn}
\affiliation{%
School of Physical Sciences, University of Chinese Academy of Sciences, Beijing 100049, China
}%

\author{Jia-Jun Wu}
\email{corresponding author: wujiajun@ucas.ac.cn}
\affiliation{School of Physical Sciences, University of Chinese Academy of Sciences, Beijing 100049, China
}%
\affiliation{%
Southern Center for Nuclear-Science Theory (SCNT), Institute of Modern Physics,
Chinese Academy of Sciences, Huizhou 516000, Guangdong Province, China
}%

\date{\today}

\begin{abstract}
As the lightest isovector vector meson, the $\rho$ meson is an important object for investigating the structure of resonant states in strong interactions. 
Owing to its strong coupling to the $\pi\pi$ channel and its large decay width, the conventional constituent quark model treatment, in which it is simply regarded as a pure $q\bar q$ bound state while the hadronic-channel coupling effects are neglected, is insufficient to fully characterize its physical properties. 
To this end, in the present work we establish a unified framework for studying the structure and resonant properties of the $\rho$ meson by combining the quark-gluon and hadronic degrees of freedom. 
At the quark-gluon level, we first determine the parameters of the chiral quark model by refitting a set of narrow mesons for which open Okubo-Zweig-Iizuka-allowed strong-decay channels are absent or strongly suppressed. 
With these parameters fixed, the bare mass of the $\rho$ meson is obtained and used as the input for the subsequent hadronic-level analysis.
At the hadronic level, based on inverse-scattering theory, we construct a model including the coupling between the bare state and the $\pi\pi$ continuum, extract the $\rho_0-\pi\pi$ interaction using the $P$-wave $\pi\pi$ scattering phase-shift data, and further calculate the width of the $\rho$ meson as well as the bare-state component in the physical state.
The present work also provides a generalizable analytical framework for further studies of other hadronic resonances with significant coupled-channel effects.
\end{abstract}

\maketitle

\section{Introduction}

Quantum chromodynamics (QCD) is the fundamental theory describing the strong interaction in which hadrons emerge as bound states of quarks and gluons. 
Because of the intrinsic property of color confinement, colored quarks and gluons cannot be isolated and observed directly, and hadronic states are the minimal experimentally observable objects in strong-interaction dynamics. 
Therefore, experimental data on hadronic reactions and decay processes constitute the primary empirical basis for the study of QCD phenomena. 
To describe these observables, one needs to construct a theoretical framework that can simultaneously encompass both the quark-gluon level and the hadronic level.

The traditional quark model greatly reduces the complexity of describing hadronic systems within the Hamiltonian framework by approximating the sea-quark fluctuations and gluons inside hadrons as constituent degrees of freedom, which are referred to as constituent quarks, thereby providing a phenomenological description of hadron structure. 
Within the quark model framework, mesons are described as quark-antiquark bound states, whereas baryons are described as three-quark systems.
The interaction potential between constituent quarks usually contains several parts: the confinement potential, the one-gluon-exchange term, and additional color-singlet potential terms.
By diagonalizing the constructed Hamiltonian matrix, one can obtain mass predictions for various hadronic states.
This is precisely the basic strategy of potential models for determining the hadron mass spectrum.
By virtue of this method, potential models have achieved many theoretical successes~\cite{PhysRevD.32.189,Isgur:1978xj,Capstick:1986ter,Wang:1992wi,Wang:1995bg,Yang:2020atz,PhysRevD.80.114023,Huang:2015uda,PhysRevD.108.054025,Vijande:2004he,Ebert:2009ub,Song:2015nia,Deng:2016stx,PhysRevD.96.116016,PhysRevD.77.074008,Liu:2019zuc,Liu:2019zoy,PhysRevD.109.116006,6x3t-x15s,Pang:2017dlw,PhysRevLett.34.369,PhysRevD.17.3090,Eichten:1979ms}.
In addition to potential models, on the basis of the quark model, various phenomenological models have been constructed to further characterize the nonperturbative effects of low-energy QCD, thereby enabling more accurate descriptions and predictions of hadron spectra and structural properties. 
For example, the MIT bag model \cite{Johnson:1975zp,Chodos:1974je,Chodos:1974pn,DeGrand:1975cf} realizes confinement by introducing quark motion in a finite spatial region together with vacuum pressure and has been successfully applied to calculations of ground-state properties and mass spectra; meanwhile, the color flux-tube model~\cite{Nambu:1974zg,BICUDO2012440,PhysRevD.31.2910}, based on a linear confinement potential, describes hadrons as composed of quark endpoints and stringlike color fields and performs well in explaining excited spectra and Regge trajectories.

However, this treatment with quarks as the fundamental degrees of freedom essentially neglects the important influence of hadronic-level interactions, especially for resonances with sizable strong-decay widths.
The coupling between such resonances and their decay channels can induce significant mass shifts, which implies that the physical states observed in experiment actually reflect a complicated coupling between quark-gluon dynamics and hadron-hadron interactions.
To better account for these effects, various theoretical approaches incorporating hadron-hadron interactions have been developed, including meson-exchange models and effective field theories.
When studying scattering and resonance problems, these methods usually require solving scattering equations (for example, the Lippmann-Schwinger equation) to obtain scattering amplitudes or the $T$-matrix, and then determining the nature of the relevant states by analyzing the pole structure in the complex energy plane. 
Comprehensive reviews of hadron-hadron interaction and scattering approaches can be found in Refs.~\cite{Machleidt1987,Myhrer1988,Machleidt:1989tm,Ecker1995,RevModPhys.77.1423,Meng2023,QUIGG1979167,LUCHA1991127}. 
In the light-meson sector, $\pi\pi$ scattering and related resonances have also been studied with $S$-matrix analyses and unitarized quark model/Friedrichs-Lee schemes, where pole structures and the dressing of bare $q\bar q$ seeds by meson continua were examined~\cite{Gao:2022dln,Zhou:2010ra,Zhou:2020vnz}.

In principle, if sufficiently precise and comprehensive scattering data are available, the extraction of resonance parameters should exhibit only weak model dependence, namely, the information extracted from different reasonable models should be consistent. 
However, due to the limited experimental data and the sensitivity to the parametrization form of the potential function, substantial theoretical uncertainties still remain in practical calculations. Therefore, the development of model-independent methods for extracting resonance properties from scattering data remains an active research direction in hadron spectroscopy. 
Following this goal, while keeping the model approximations required in the present analysis, we combine the constituent quark model with the hadronic-level inverse-scattering method of Ref.~\cite{Li:2022aru} to study the $\rho$ meson.

Our approach contains the following key steps.
First, we adopt the chiral quark model (ChQM), which contains the confinement potential, one-gluon exchange, and meson-exchange terms derived from chiral effective field theory. 
The model parameters are determined by fitting the ground-state pseudoscalar mesons, vector mesons, and some narrow $P$-wave mesons. 
Meanwhile, the $\rho$, $K^*$, and $D^*$ are not included in the fit, because their masses can be affected by the coupling to the corresponding Okubo-Zweig-Iizuka-allowed two-pseudoscalar channels. 
In this way, the model parameters are mainly constrained by states for which such coupled-channel effects are less prominent.
After the fitting, the bare-state mass of the $\rho$ meson is found to be about $845$ MeV, which is significantly higher than the experimentally observed Breit-Wigner central mass of $770$ MeV. 

In the second step, we study the $\rho_0-\pi\pi$ system, where $\rho_0$ denotes the bare state. 
We consider a scenario (which we call the $bc$ model) that contains only the interaction between the bare state $\rho_0$ and the $\pi\pi$ channel. 
The key point is that no specific functional form is assumed a priori; instead, the interaction form factor is directly reconstructed from the $P$-wave $\pi\pi$ scattering phase shift together with the bare-mass constraint from the quark model. 
This method not only constrains the range of the interaction strength, but also predicts the decay width at the point where the phase shift satisfies $\delta = 90^{\circ}$, as well as the overlap between the physical $\rho$ state and the bare state. 

The main innovation of this work lies in the unified treatment of the quark-gluon and hadronic degrees of freedom and in the extraction of the form factor by inverse-scattering techniques without relying on phenomenological assumptions. 

The remainder of this paper is organized as follows. 
In Sec.~\ref{sec:ch_quark}, we introduce the formalism of the quark model and the determination of its parameters. 
Section~\ref{sec:Tmatrix} presents the theoretical framework of the inverse-scattering analysis. 
Section~\ref{sec:results} shows the numerical results and discussion. 
Finally, a summary and outlook are given in Sec.~\ref{sec:sum}.

\section{The chiral quark model}
\label{sec:ch_quark}

\subsection{Model framework}

The quark model calculation is used to determine the bare input for the subsequent hadronic-level analysis. 
In particular, we refit the model parameters and extract the bare mass of the $\rho$ meson within ChQM. 
Following the formulation of Vijande \textit{et al}.~\cite{Vijande:2004he}, the Hamiltonian is written as
\begin{align}
H=&\sum_i\left(m_i+\frac{p_i^2}{2m_i}\right)-T_{CM}\notag\\
&+V_{\mathrm{CON}}(\vec{r}_{ij})+V_{\mathrm{OGE}}(\vec{r}_{ij})+V_{\chi}(\vec{r}_{ij}),
\end{align}
where $m_i$ and $p_i$ denote the mass and momentum of the $i$th quark, respectively, and $T_{CM}$ represents the kinetic energy of the center of mass of the system.
$V_{\mathrm{CON}}(\vec{r}_{ij})$, $V_{\mathrm{OGE}}(\vec{r}_{ij})$, and $V_{\chi}(\vec{r}_{ij})$ denote the confinement potential, the one-gluon-exchange (OGE) potential, and the meson-exchange potential, respectively, with $\chi = \pi, K, \eta, \sigma$.

The OGE potential can be decomposed as $V_{\mathrm{OGE}} = V_{\mathrm{OGE}}^{C} + V_{\mathrm{OGE}}^{T} + V_{\mathrm{OGE}}^{SO}$, where the first term is the central potential, the second term represents the tensor interaction, and the last term corresponds to the spin-orbit interaction. 
Their explicit forms are given by
\begin{align}
&V_{\mathrm{OGE}}^{C}(\vec{r}_{ij})=\frac{\alpha_s}{4}\vec\lambda_i^{c}\cdot\vec\lambda_j^{c}
\left[\frac{1}{r_{ij}}-\frac{1}{6m_im_j}\vec{\sigma}_i\cdot\vec{\sigma}_j \frac{e^{-r_{ij}/r_0(\mu)}}{r_{ij}r_0^2(\mu)}
    \right],
\end{align}
\begin{align}
&V_{\mathrm{OGE}}^T(\vec r_{ij})
=-\frac{\alpha_s}{16m_im_j}\vec\lambda_i^{c}\cdot\vec\lambda_j^{c}\,S_{ij}
\notag
\\
&\quad\quad\,\,\,\times\Biggl[\frac{1}{r_{ij}^3}
-\frac{e^{-r_{ij}/r_g(\mu)}}{r_{ij}}
\Bigl(\frac{1}{r_{ij}^2}
+\frac{1}{3r_g^2(\mu)}
+\frac{1}{r_{ij}r_g(\mu)}\Bigr)
\Biggr]\,,
\\
&V_{\mathrm{OGE}}^{SO}(\vec r_{ij})=-\frac{\alpha_s}{16m_i^2m_j^2}\vec\lambda_i^{c}\cdot\vec\lambda_j^{c} \ V_1\notag\\
&\quad\quad\,\,\,\times\Biggl[\frac{1}{r_{ij}^3}
-\frac{e^{-r_{ij}/r_g(\mu)}}{r_{ij}^3}\Bigl(1+\frac{r_{ij}}{r_g(\mu)}\Bigr)
\Biggr]\ ,
\end{align}
where $r_{ij}=|\vec{r}_{ij}|$, $r_0(\mu)=\hat r_0/\mu$, $r_g(\mu)=\hat r_g/\mu$, and $\mu$ is the reduced mass of the quark-antiquark pair.
Here, $\hat{r}_0$ and $\hat{r}_g$ are fitting parameters related to the replacement of $\delta(r_{ij})$ by smeared functions~\cite{Weinstein:1982gc,Weinstein:1983gd,Weinstein:1990gu,Bhaduri:1980fd}.
The tensor operator is defined as
\[
S_{ij}=3(\vec{\sigma}_i\cdot\vec{r}_{ij})(\vec{\sigma}_j\cdot\vec{r}_{ij})/r_{ij}^2-\vec{\sigma}_i\cdot\vec{\sigma}_j,
\]
where $\vec{\sigma}_i$ denotes the Pauli operator of the $i$th quark.
The expression for $V_1$ is
\[
V_1=
\bigl((m_i+m_j)^2+2m_im_j\bigr)\vec S_{+}\cdot\vec L
+\bigl(m_j^2 - m_i^2\bigr)\vec S_{-}\cdot\vec L,
\]
where $\vec{S}_{\pm}=\vec S_i\pm \vec S_j$, $\vec S_i$ is the spin of the $i$th quark, and $\vec L$ is the orbital angular momentum of the two-quark system.
The matrix $\lambda_i^c$ denotes the $SU(3)$ color Gell-Mann matrix of the $i$th quark.
For a quark-antiquark pair, the term $\vec\lambda_i^c\cdot\vec\lambda_j^c$ should be replaced by $-\vec\lambda_i^c\cdot\vec\lambda_j^{c*}$.
This minus sign may originate from the different Wick contractions of the vector boson (gluon) in the lowest-order Feynman diagrams for quark-quark and quark-antiquark pairs, while the transpose or complex conjugation arises from the rule that the fermion line of the antiquark is opposite in direction to that of the quark.

The confinement potential can be decomposed as $V_{\mathrm{CON}}=V_{\mathrm{CON}}^{C}+V_{\mathrm{CON}}^{SO}$, with explicit forms
\begin{align}
V_{\mathrm{CON}}^{C}(\vec{r}_{ij})&=\left[-a_c(1-e^{-\mu_c r_{ij}})+ \Delta
\right]\bigl(\vec\lambda_i^{c}\cdot\vec\lambda_j^{c}\bigr),
\label{con}
\\
V_{\mathrm{CON}}^{SO}(\vec r_{ij})
&=-\bigl(\vec\lambda_i^{c}\cdot\vec\lambda_j^{c}\bigr)
\frac{a_c\,\mu_c\,e^{-\mu_c\,r_{ij}}}{4 m_i^2 m_j^2 r_{ij}}\ V_2,
\end{align}
where $\Delta$ is a global constant fixing the origin of energies, and
\begin{align}
V_2=&
\bigl((m_i^2 + m_j^2)(1-2a_s) 
+4m_im_j(1 - a_s)\bigr)\vec S_{+}\cdot\vec L
\notag\\
&+ (m_j^2 - m_i^2)(1-2a_s)\vec S_{-}\cdot\vec L.
\end{align}
It should be noted that although the operator $S_{-}$ in $V_1$ and $V_2$ leads to spin mixing, its contribution is rather small and is therefore neglected in the present work.
Similarly, $S_{ij}$ also induces orbital-angular-momentum mixing, but its effect is likewise small in our calculations, and thus the off-diagonal matrix elements generated by this mixing are also neglected directly.

The meson-exchange potential is written as
\[
V_{\chi}=V^C_{\chi}+V_{\chi}^{T}+V_{\chi}^{SO},
\]
where
\begin{align}
V_{\chi}^{C}(\vec{r}_{ij}) &=\frac{g^2_{ch}}{4\pi} \frac{m_{\chi}^3}{12m_i m_j} \frac{\Lambda^2_{\chi}}{\Lambda^2_{\chi} - m_{\chi}^2} \notag\\
&\times\left[Y(m_{\chi}r_{ij}) - \frac{\Lambda_{\chi}^3}{m_{\chi}^3} Y(\Lambda_{\chi}r_{ij})\right](\vec{\sigma}_i \cdot \vec{\sigma}_j)\ W_\chi,
\\
V_{\chi}^{T}(\vec{r}_{ij}) &= \frac{g_{\rm ch}^2}{4\pi} \frac{m_{\chi}^2}{12\,m_i\,m_j} \frac{\Lambda_{\chi}^2}{\Lambda_{\chi}^2 - m_{\chi}^2} m_{\chi} 
\notag\\
&\times\left[ H(m_{\chi} r_{ij}) - \frac{\Lambda_{\chi}^3}{m_{\chi}^3} H(\Lambda_{\chi} r_{ij}) \right] S_{ij} \ W_\chi,
\end{align}
where $Y(x)=e^{-x}/x$, $H(x)=(1+3/x+3/x^2)Y(x)$.
The expression for $W_\chi$ is
\begin{align}
W_\pi &= \sum_{a=1}^3 (\lambda_i^a \lambda_j^a),
\\
W_K &= \sum_{a=4}^7 (\lambda_i^a \lambda_j^a),
\\
W_{\eta}&=\left(\cos\theta_P\  \lambda^8_i \cdot \lambda^8_j - \sin\theta_P \ \lambda^0_i \cdot \lambda^0_j\right).
\end{align}
For a quark ($i$)-antiquark ($j$) system, one has $\lambda_i^{a}\cdot\lambda_j^{a}\rightarrow\lambda_i^{a}\cdot\lambda_j^{a*}$.
For the scalar meson $\sigma$, the corresponding potential is written as
\begin{align} 
V_{\sigma}^{C}(\vec{r}_{ij})&=-\frac{g^2_{ch}}{4\pi}\frac{\Lambda_{\sigma}^2}{\Lambda^2_{\sigma}-m_{\sigma}^2}m_{\sigma}\notag\\
&\times\left[Y(m_{\sigma}r_{ij})
-\frac{\Lambda_{\sigma}}{m_{\sigma}}Y(\Lambda_{\sigma}r_{ij})\right],
\\
V_\sigma^{SO}(\vec r_{ij})
&=-\frac{g_{\rm ch}^2}{4\pi}\,
\frac{\Lambda_\sigma^2}{\Lambda_\sigma^2 - m_\sigma^2}\,
\frac{m_\sigma^3}{2\,m_i\,m_j}\notag\\
&\times\Bigl[
G(m_\sigma r_{ij})
-\frac{\Lambda_\sigma^3}{m_\sigma^3}\,
G(\Lambda_\sigma r_{ij})
\Bigr]
\vec L\cdot\vec S,
\end{align}
where
\[
G(x)=\left(1+\frac{1}{x}\right)\frac{Y(x)}{x},
\]
and $\vec{S}$ denotes the total spin of quarks $i$ and $j$.

\subsection{Running coupling constant}

In the ChQM, the low-energy effects of gluons and sea-quark fluctuations are effectively absorbed into constituent quark masses and effective interactions. 
This treatment reduces the number of active degrees of freedom (d.o.f.) and allows the hadron spectrum to be calculated within a few-body Hamiltonian framework. 
However, it also changes the meaning of the running coupling constant $\alpha_s$. 
In QCD, $\alpha_s$ describes the interaction among current quarks and gluons, whereas in the present model it appears as an effective coupling between constituent quarks in the one-gluon-exchange potential. 
Thus, its scale dependence has to be parametrized phenomenologically.

Inspired by the functional form of the $\alpha_s$ curve in Refs.~\cite{Vijande:2004he,PhysRevD.32.189}, we classify all mesons into ten different configurations according to the flavor structure of the constituent quarks: $qq$ ($q=u,d$), $sq$, $ss$, $cq$, $cs$, $cc$, $bq$, $bs$, $bc$, and $bb$. 
Correspondingly, there are ten different reduced masses $\mu$ serving as the energy scale.
For these ten cases, we introduce ten mutually independent $\alpha_s$ parameters to fit the hadron spectrum, while the remaining free parameters are shared among all configurations.
This procedure effectively introduces ten extra parameters for the running coupling constant in the one-gluon-exchange potential, whereas standard lowest-order perturbative QCD usually involves only one parameter (namely $\Lambda$), as shown in Ref.~\cite{PhysRevD.110.030001}.
The resulting behavior is shown in Fig.~1(a), from which one can clearly see that $\alpha_s$ decreases as the reduced mass $\mu$ increases, exhibiting a logarithmically suppressed behavior consistent with the QCD expectation.
It is worth noting that, for the three cases $sq$, $cq$, and $bq$, the values of $\alpha_s$ are very close to one another, and their reduced masses are also approximately close to the mass of the light constituent quark.
However, there are still obvious deviations on a logarithmic scale.
These results indicate that there may be two possible directions for improvement:
First, the definition of the energy scale may need to go beyond the simple reduced-mass form and include additional correction terms;
second, the functional form of the logarithmic dependence itself may also require further improvement.
Based on these considerations, we construct the following three alternative models.

First, we redefine a new variable $Q$ to replace $\mu$,
\begin{equation}
Q=\mu\left(1+e^{-(m_i^2+m_j^2)/c_1^2}+e^{-(\mu-c_2)^2/c_2^2}\right).
\end{equation}
The two additional parameters $c_1$ and $c_2$ are used to adjust the relation between the energy scale and the two-quark masses.
The term associated with $c_1$ raises the energy scale for light-quark systems, whereas the term associated with $c_2$ shifts all energy scales toward $c_2$; the fitting result shows that $c_2$ is located roughly near the $c$-quark mass.
On this basis, we consider modifying the logarithmic dependence of $\alpha_s$ to describe its scale behavior more flexibly.
In Model~1, the power behavior of the logarithm is changed as
\begin{equation}
    \alpha_s(Q)=\frac{\alpha_0}{\ln^k(Q^2/\Lambda_0^2)},
\end{equation}
where the parameter $k$ is used to adjust the decay rate of the coupling constant with the energy scale.
In Model~2, an infrared correction term is introduced,
\begin{equation}
    \alpha_s(Q)=\frac{\alpha_0}{\ln((Q^2-\mu_0^2)/\Lambda_0^2)},
\end{equation}
This form shifts the effective starting point of the logarithmic running by introducing an additional low-energy scale $\mu_0$, thereby modifying the behavior of the coupling constant in the medium- and low-energy regions. Compared with the standard one-loop form, this parametrization retains the logarithmic falloff in the high-energy region while allowing a more flexible variation of the coupling constant in the actual fitting energy region.
In Model~3, higher-order corrections are simulated through
\begin{equation}
    \alpha_s(Q)=\frac{\alpha_0}{\ln(Q^2/\Lambda_0^2)}-\frac{\alpha_1\ln(\ln(Q^2/\Lambda_0^2))}{\ln^2(Q^2/\Lambda_0^2)},
\end{equation}
This expression is inspired by the next-to-leading-order expansion form of $\alpha_s$ in QCD and is used to describe the running behavior of the coupling constant more accurately.

\begin{figure*}[t]
    \centering
    \includegraphics[width=\textwidth]{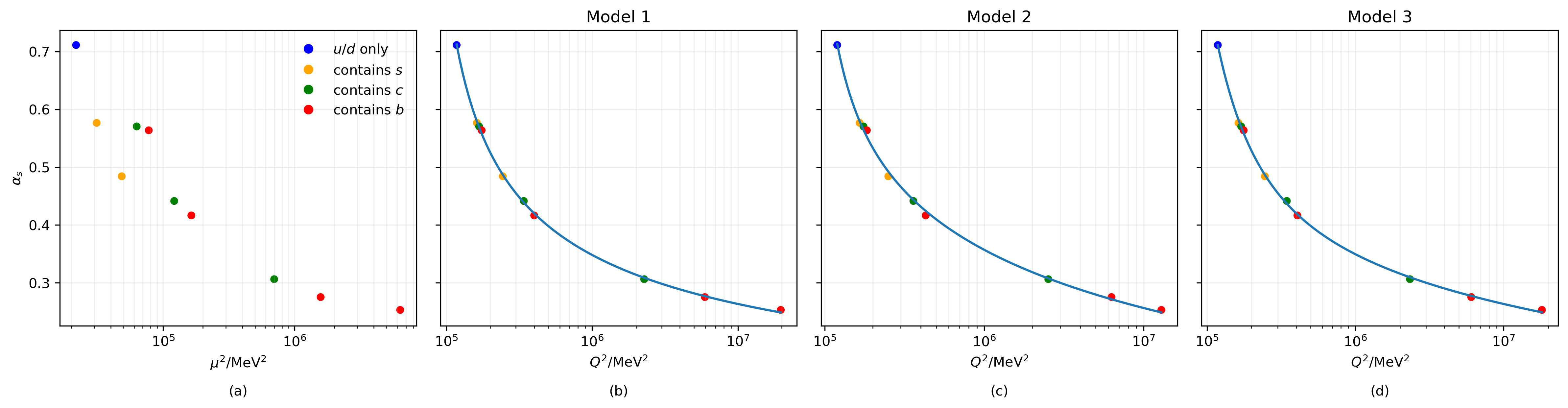}
    \label{fig:alphas_scatterplot}
    \caption{Running coupling constant. 
    Panel (a) shows the $\alpha_s$ values extracted independently from the different cases: the blue points for $\alpha_{qq}(q=u,d)$, the orange points for $\alpha_{s(q/s)}$, the green points for $\alpha_{c(q/s/c)}$, and the red points for  $\alpha_{b(q/s/c/b)}$. 
    Panels (b)--(d) show the results from Models~1--3, respectively.}
\end{figure*}

\subsection{Parameter fitting and bare-mass prediction}\label{sec:para}

In this work, we employ the Gaussian expansion method (GEM) \cite{Hiyama:2003cu} to handle the calculation of the spatial wave functions involved in the above quark model. 
The spatial wave function is expressed as
\begin{equation}
    	\Psi_{lm}(\vec{r})=\sum_{n=1}^{N}c_{nl}\phi^G_{nlm}({\vec{r}}).
\end{equation}
The Gaussian basis function is
\begin{equation}
    \phi^G_{nlm}({\vec{r}})=\left(\frac{2^{l+2}(2v_n)^{l+\frac{3}{2}}}{\sqrt{\pi}(2l+1)!!} \right)^{\frac{1}{2}}r^l e^{-v_n r^2} Y_{lm}(\hat{r}).
\end{equation}
Here $l$ and $m$ are related to the total angular momentum $J$ and its third component $J_z$.
%
$N$ denotes the number of Gaussian basis functions.
The Gaussian width parameters $v_n$ ($n=1,\ldots,N$) are determined by
\begin{equation}
    v_n=\frac{1}{r_1^2}\left(\frac{r_1}{r_N}
    \right)^{\frac{n-1}{N-1}}.
\end{equation}

The eigenenergies and the Gaussian expansion coefficients can be obtained through the Rayleigh-Ritz variational principle in the form
\begin{equation}
    \sum_{n'=1}^{N}\big[(T_{nn'}+V_{nn'})-EN_{nn'}\big]c_{n'l}=0,\label{eq:eigen}
\end{equation}
where $T$ is the kinetic-energy matrix and $V$ is the potential-energy matrix, containing the three kinds of interaction potentials introduced in the previous subsection.
As $N$ increases, the stability of the numerical results during the iteration process must be ensured.

By solving Eq.~(\ref{eq:eigen}), one can obtain the meson spectrum.
In this work, we choose 29 very narrow mesons, as listed in Table~\ref{tab:fitresult}.
It should be pointed out that the three vector mesons $\rho$, $K^*$, and $D^*$ all have open strong-interaction decay channels and are therefore not suitable as fitting inputs.
As mentioned above, each potential model contains 18 free parameters (see Table~\ref{tab:fit_parameter}), which are determined by fitting the above 29 mesons.
The $\chi^2$ function adopted for our fitting procedure is defined as 
\begin{equation}
\chi^2 = \sum_i \frac{(m_i^{\rm cal} - m_i^{\rm exp})^2}{\sigma^2},
\end{equation}
where the target error is taken as $\sigma = 5$~MeV. 
With this choice, the three fits give $\chi^2 \simeq 16.9$ and $\chi^2/\mathrm{d.o.f.} \simeq 1.5$ for model~1, $\chi^2 \simeq 20.3$ and $\chi^2/\mathrm{d.o.f.} \simeq 1.8$ for model~2, and $\chi^2 \simeq 17.5$ and $\chi^2/\mathrm{d.o.f.} \simeq 1.6$ for model~3.
On this basis, we predict the bare masses of $\rho$, $K^*$, and $D^*$.
The fitted meson spectra and the corresponding parameters are listed in Table~\ref{tab:fitresult} and Table~\ref{tab:fit_parameter}, respectively.
The calculated masses of the 29 fitted mesons from the three models are in good agreement with the experimental data. For most states, the mass deviations are within $5$~MeV.

Once the quark model parameters are determined from the 29 selected mesons, the model yields the bare-mass predictions for the excluded $\rho$, $K^*$, and $D^*$ states. 
These bare masses must be distinguished from the measured physical masses, which already incorporate open-channel dressing effects. 
The magnitude of this dressing depends on both the coupling strength to the relevant channel and the proximity of the state to the corresponding threshold. 
When sufficient phase space is available, the coupling strength is directly reflected in the decay width; consequently, a broad state can receive a sizable mass shift even if the relevant threshold is located far below its physical mass. 
Conversely, a very small decay width does not necessarily mean that the channel coupling is weak, since the width can be strongly suppressed by limited phase space when the state lies extremely close to the threshold. 
The $D^{*}(2007)^{0}$ provides a typical example of this latter situation. 
Its mass lies just above the $D\pi$ threshold, so the available phase space is strongly suppressed. 
This threshold proximity explains its extremely small width, but it does not imply a weak coupling to the $D\pi$ channel. 
Because this coupling remains strong, it still generates a sizable downward mass shift relative to the bare mass predicted by the quark model~\cite{PhysRevD.109.116006}.
For this reason, the $D^*$ states are excluded from the global fit to ensure a reliable extraction of the quark model parameters.

For the $\rho$ meson, the necessary distinction between bare and physical masses is particularly evident when examining its relation to the $\omega$ meson. 
Although the physical $\rho$ and $\omega$ mesons are nearly degenerate in mass, they exhibit vastly different decay widths: the $\rho$ meson has a width of about $140$~MeV, whereas the $\omega$ meson has a width of only a few MeV. 
Because the broad $\rho$ meson receives a much larger coupled-channel mass shift than the narrow $\omega$ meson, one expects their bare masses to be separated appreciably before these channel contributions are taken into account. 
In the present potential model, this required bare-mass splitting is naturally generated by the chiral potential $V_{\pi}$, which differentiates the two states through their different isospin and flavor wave functions. 
If one artificially forces the bare $\rho$ mass to remain close to the physical $\omega$ mass, this physical constraint is violated, and a simultaneous description of the remaining fitted mesons becomes difficult. 
Therefore, the near-degeneracy observed in the physical $\rho$ and $\omega$ masses should instead be recovered only after the coupled-channel effects are included.

The bare-mass assignment can be checked more directly by restoring $\rho$, $K^*$, and $D^*$ in the global fit.
These states are not used as inputs in the main fit, since their measured masses contain shifts from open channels, while the Hamiltonian in Sec.~\ref{sec:ch_quark} gives bare quark model eigenvalues.
With the three states restored, we make two fits which differ only in the masses used for them. 
In fit~1, the three input masses are fixed to the bare masses obtained in the main fit. 
The comparison with the main fit checks whether these added bare inputs change the fitted parameters and the remaining fitted masses. 
In fit~2, the three input masses are replaced by their measured physical masses. 
This fit tests what is obtained if the measured masses, which already include the open-channel shifts, are used as inputs of the same bare Hamiltonian. 
For simplicity, this check is carried out with model~1.
Models~1--3 differ only in the parametrization of $\alpha_s$ within the same potential model, and the same qualitative behavior is expected for models~2 and 3. 
The fitted parameters and spectra are given in Appendix~\ref{app:fit-stability}. 
Fit~1 stays close to the main fit, whereas fit~2 drives the remaining narrow spectrum away from experiment. 
The stability of fit~1 and the large deviation found in fit~2 support treating the quark model eigenvalues of $\rho$, $K^*$, and $D^*$ as bare masses.

In summary, at the quark-gluon level, we employ the chiral quark model to fit narrow-width mesons, thereby obtaining the bare mass of the $\rho$ meson.
In the method of the present work, such a bare state contains information from the quark-gluon level and will be used as a basic building block in the subsequent hadronic-level study.

\begin{table}[htpb]
\caption{
Quark model parameters for models 1--3. 
The masses of $\pi$, $K$, and $\eta$ are fixed at their experimental values, and $\Lambda_{\sigma}$, $\Lambda_{\pi}$, $\Lambda_{K}$, and $\Lambda_{\eta}$ are taken from Ref.~\cite{Vijande:2004he}. 
For each model, the remaining 18 parameters are fitted to the selected meson spectrum. 
The input pseudoscalar mixing angle $\theta_p$ and the calculated $\eta_1-\eta_8$ mixing angle $\theta_p^{\rm cal}$ are listed together; see Appendix~\ref{MixingAngle} for details.
}
\label{tab:fit_parameter}
\begin{ruledtabular}
\begin{tabular}{lrrr}
& Model 1 & Model 2 & Model 3\\

\textbf{Quark masses} & & & \\
$m_u = m_d$ (MeV) & 295 & 295 & 295 \\
$m_s$ (MeV) & 440 & 440 & 440 \\
$m_c$ (MeV) & 1674.4 & 1658 & 1674.5\\ 
$m_b$ (MeV) & 5035.4 & 5022 & 5039.6 \\\\

\hline \\

\textbf{Confinement} & & &  \\
$a_c$ (MeV) & 326.1 & 322.8 & 327.2\\
$\mu_c$ (fm$^{-1}$) & 1.258 & 1.290 & 1.256\\
$\Delta$ (MeV) & 170.6 & 166.8 & 172.5 \\
$a_s$  & 0.715 & 0.72 & 0.714 \\
\\ \hline \\

\textbf{Goldstone bosons} & & & \\
$m_{\pi}$ (fm$^{-1}$) &   & 0.70 (fixed) &  \\
$m_{\eta}$ (fm$^{-1}$) & & 2.77 (fixed) & \\
$m_{K}$ (fm$^{-1}$) & & 2.51 (fixed) & \\
$\Lambda_{\pi}=\Lambda_{\sigma}$ (fm$^{-1}$) & & 4.20 (fixed) & \\
$\Lambda_{\eta}=\Lambda_{K}$ (fm$^{-1}$) & & 5.20 (fixed) & \\
$m_{\sigma}$ (fm$^{-1}$) & 3.594 & 2.60 & 4.15 \\
$g^2_{ch}/(4\pi)$ & 0.4489 & 0.4451 & 0.4452 \\
$\theta_P$ ($^\circ$) & -20.6 & -20 & -21 \\
$\theta_P$($^\circ$)-calculated & -20.59 & -20.06 & -20.96 \\
\\ \hline \\

\textbf{OGE} & & & \\
$\hat{r}_0$ (MeV fm) & 35.739 & 37.480 & 35.284 \\
$\hat{r}_g$ (MeV fm) & 25.298 & 26.166 & 25.081 \\
$\alpha_0$ & 0.5634 & 2.093 & 3 \\
$\Lambda_0$ (MeV) & 258.9 & 56.12 & 128.1 \\
$c_1$  (MeV) & 1233 & 1222 & 1238 \\
$c_2$ (MeV) & 1402 & 1248 & 1365 \\
$\alpha_1$ & $\cdots$ & $\cdots$ & 4.54 \\
$\mu_0$ (MeV) & $\cdots$ & 265 & $\cdots$ \\
$k$  & 0.4689 & $\cdots$ & $\cdots$ \\ 

\end{tabular}
\end{ruledtabular}
\end{table}

\begin{table*}[htpb]
\caption{
Meson spectra obtained with models 1--3, compared with the Godfrey-Isgur (GI) model~\cite{PhysRevD.32.189} and experimental data~\cite{PhysRevD.110.030001}. 
The 18 model parameters are fitted to the experimental masses of the 29 states, with $\rho$, $K^*$, and $D^*$ excluded from the fit.
}
\label{tab:fitresult}
\begin{ruledtabular}
\begin{tabular}{@{}cccccccc@{}}
$n\,{}^{2S+1}L_J$ & $J^{P(C)}$ & State & Model 1 & Model 2 & Model 3 & GI model~\cite{PhysRevD.32.189} & Expt.~\cite{PhysRevD.110.030001} \\
\colrule

$1\,{}^1S_0$ & $0^{-+}$ & $\pi$        & 135.7   & 136.4   & 136.1   & 150   & 135   \\
$1\,{}^1S_0$ & $0^{-+}$ & $\eta$       & 547.4   & 545.7   & 548.0   & 520   & 548   \\
$1\,{}^3S_1$ & $1^{--}$ & $\rho$       & 846.8   & 848.9   & 845.5   & 770   & 775   \\
$1\,{}^3S_1$ & $1^{--}$ & $\omega$     & 779.6   & 777.8   & 779.8   & 780   & 783   \\
$1\,{}^1S_0$ & $0^{-+}$ & $\eta'$      & 957.5   & 958.8   & 956.9   & 960   & 958   \\
$1\,{}^3S_1$ & $1^{--}$ & $\phi$       & 1022.7  & 1023.8  & 1022.8  & 1020  & 1019  \\
$1\,{}^1S_0$ & $0^{-}$  & $K$          & 497.0   & 497.4   & 495.8   & 470   & 498   \\
$1\,{}^3S_1$ & $1^{-}$  & $K^*$        & 950.4   & 954.1   & 948.6   & 900   & 896   \\
$1\,{}^1S_0$ & $0^{-}$  & $D$          & 1863.5  & 1864.6  & 1863.1  & 1880  & 1865  \\
$1\,{}^3S_1$ & $1^{-}$  & $D^*$        & 2027.2  & 2030.0  & 2026.1  & 2040  & 2007  \\
$1\,{}^1S_0$ & $0^{-}$  & $D_s$        & 1971.5  & 1968.5  & 1973.6  & 1980  & 1968  \\
$1\,{}^3S_1$ & $1^{-}$  & $D_s^*$      & 2112.3  & 2114.2  & 2112.2  & 2130  & 2112  \\
$1\,{}^1S_0$ & $0^{-+}$ & $\eta_c$     & 2984.1  & 2982.4  & 2983.7  & 2970  & 2984  \\
$1\,{}^3S_1$ & $1^{--}$ & $J/\psi$     & 3096.8  & 3095.2  & 3096.6  & 3100  & 3097  \\
$1\,{}^3P_0$ & $0^{++}$ & $\chi_{c0}$  & 3414.8  & 3414.8  & 3414.8  & 3440  & 3415  \\
$1\,{}^3P_1$ & $1^{++}$ & $\chi_{c1}$  & 3509.6  & 3510.3  & 3509.4  & 3510  & 3511  \\
$1\,{}^1P_1$ & $1^{+-}$ & $h_c$        & 3528.7  & 3529.3  & 3528.6  & 3520  & 3525  \\
$1\,{}^3P_2$ & $2^{++}$ & $\chi_{c2}$  & 3553.8  & 3554.7  & 3553.7  & 3550  & 3556  \\
$2\,{}^3S_1$ & $1^{--}$ & $\psi(2S)$   & 3684.9  & 3683.7  & 3685.3  & 3680  & 3686  \\
$1\,{}^1S_0$ & $0^{-+}$ & $\eta_b$     & 9394.9  & 9395.7  & 9394.9  & 9400  & 9399  \\
$1\,{}^3S_1$ & $1^{--}$ & $\Upsilon$   & 9465.2  & 9464.0  & 9465.3  & 9460  & 9460  \\
$1\,{}^3P_0$ & $0^{++}$ & $\chi_{b0}$  & 9858.6  & 9858.7  & 9858.6  & 9850  & 9859  \\
$1\,{}^3P_1$ & $1^{++}$ & $\chi_{b1}$  & 9890.3  & 9890.0  & 9890.2  & 9880  & 9893  \\
$1\,{}^1P_1$ & $1^{+-}$ & $h_b$        & 9898.6  & 9898.2  & 9898.5  & 9880  & 9899  \\
$1\,{}^3P_2$ & $2^{++}$ & $\chi_{b2}$  & 9909.5  & 9909.0  & 9909.4  & 9900  & 9912  \\
$2\,{}^3S_1$ & $1^{--}$ & $\Upsilon(2S)$ & 10035.3 & 10034.8 & 10035.6 & 10000 & 10023 \\
$1\,{}^1S_0$ & $0^{-}$  & $B$          & 5275.7  & 5276.4  & 5275.4  & 5310  & 5280  \\
$1\,{}^3S_1$ & $1^{-}$  & $B^*$        & 5333.5  & 5334.8  & 5332.8  & 5370  & 5325  \\
$1\,{}^1S_0$ & $0^{-}$  & $B_s$        & 5362.7  & 5360.5  & 5363.7  & 5390  & 5367  \\
$1\,{}^3S_1$ & $1^{-}$  & $B_s^*$      & 5413.8  & 5413.5  & 5414.0  & 5450  & 5415  \\
$1\,{}^1S_0$ & $0^{-}$  & $B_c$        & 6275.1  & 6279.7  & 6274.9  & 6270  & 6274  \\
$1\,{}^3D_2$ & $2^{--}$ & $\Upsilon_2$ & 10157.8 & 10157.6 & 10158.0 & 10150 & 10164 \\
\end{tabular}
\end{ruledtabular}
\end{table*}

\section{INVERSE SCATTERING THEORY FRAMEWORK}
\label{sec:Tmatrix}

\subsection{$\rho_0-\pi\pi$ Hamiltonian}

In this section, we introduce how to determine the interaction from experimental data. 
For convenience, we mainly illustrate the formalism using the $\rho$-meson system. 
The $\rho$-meson system contains a bare $\rho$ state, denoted by $\ket{\rho_0^m}$, and its dominant decay channel $\pi\pi$, denoted by $\ket{\vec{p}}$, where $\vec{p}$ is the three-momentum of each $\pi$ in the center-of-mass frame.
The Hamiltonian of this system can be written as $\hat{H}=\hat{H}_0+\hat{H}_I$, where
\begin{equation}
\hat{H}_0=\sum_{m}\ket{\rho_0^m}m_0\bra{\rho_0^m}+\int\frac{d\vec{p}}{(2\pi)^3}h_{\vec{p}}\ket{\vec{p}}\bra{\vec{p}},
\end{equation}
where $m_0$ is the mass of the bare state $\ket{\rho_0}$, and $h_{\vec{p}}=2\sqrt{m_\pi^2+\vec{p}^2}$ is the free energy of the $\pi\pi$ channel with momentum $\vec{p}$.
The interaction Hamiltonian $\hat{H}_I$ is written as $\hat{g}+\hat{v}$,
\begin{equation}
\hat{g}=\sum_{m}\int\frac{d\vec{p}}{(2\pi)^3}\ket{\rho_0^m}g(\vec{p},m)\bra{\vec{p}}+\mathrm{H.c.},
\end{equation}
and
\begin{equation}
\hat{v}=\int\frac{d\vec{p}\,d\vec{q}}{(2\pi)^6}
\ket{\vec{p}}
v_{\alpha,\beta}(\vec{p},\vec{q})
\bra{\vec{q}}.
\end{equation}

Since the $\rho$ meson couples to the $\pi\pi$ continuum dominantly through the $P$-wave, we separate the angular dependence of the continuum state. 
The partial-wave basis is defined as
\begin{equation}
\ket{p;l,m}=\int d\Omega_{\hat{p}}Y_{lm}(\hat{p})\ket{\vec{p}},
\end{equation}
with the inverse relation
\begin{equation}
\ket{\vec{p}}=\sum_{l,m}Y^*_{lm}(\hat{p})\ket{p;l,m}.
\end{equation}
With the above normalization, one has
\begin{align}
&\langle p; l,m | q; l',m' \rangle
=\frac{(2\pi)^3}{p^2}\delta(p-q)\delta_{ll'}\delta_{mm'},\\
&\sum_m\ket{\rho_0^m}\bra{\rho_0^m}
+\sum_{l,m}\int\frac{q^2dq}{(2\pi)^3}
\ket{q;l,m}\bra{q;l,m}=1.
\end{align}

On this basis, the interaction terms can be written as
\begin{align}
\left\langle p;l,m\big|\hat{g}\big|\rho_0^{m'}\right\rangle
&=\int\frac{d\vec{q}}{(2\pi)^3}\bra{p;l,m}\vec{q}\,\rangle g(\vec{q},m')\notag\\
&=\int Y^*_{lm}(\hat{p})d\Omega_{\hat{p}}\,g(\vec{p},m')\notag\\
&\equiv \lambda_b f(p) \delta_{mm'}\delta_{l1},
\end{align}
and
\begin{align}
\left\langle p;l,m|\hat{v}|p';l',m'\right\rangle&=
\int d\Omega_{\hat{p}}d\Omega_{\hat{p}'}Y^*_{lm}(\hat{p})
Y_{l'm'}(\hat{p}')v(\vec{p},\vec{q})\notag\\
&\equiv \lambda_c \tilde{v}(p,q)\delta_{ll'}\delta_{mm'}.
\end{align}
It should be noted that the $\delta$ functions in $l$ and $m$ arise here from angular-momentum conservation.
For simplicity, in the following we abbreviate $\ket{\rho_0^m}$ and $\ket{p;l,m}$ as $\ket{\rho_0}$ and $\ket{p}$, respectively.

After the partial-wave expansion, the Hamiltonian can be written as
\begin{align}
H&=m_0\ket{\rho_0}\bra{\rho_0}+\int\frac{p^2dp}{(2\pi)^3}h_p\ket{p}\bra{p}\notag\\
&+\lambda_b\int\frac{p^2dp}{(2\pi)^3}f(p)\left(\ket{\rho_0}\bra{p}
+\ket{p}\bra{\rho_0}\right)\notag\\
&+\lambda_c\int\frac{p^2dp}{(2\pi)^3}\frac{q^2dq}{(2\pi)^3}v(p,q)\ket{p}\bra{q}.
\end{align}
If one further assumes a separable potential, one may take $v(p,q)\equiv f(p)f(q)$.

In Ref.~\cite{Li:2022aru}, a formal method was established based on the Fredholm determinant to extract the bare mass $m_0$, $(\lambda_b f(p))^2$, and $\lambda_c v(p,q)$ from the phase shift.
Within that method, under certain assumptions, the interaction can be reconstructed directly from the phase shift in real momentum space without assuming the specific forms of $f(p)$ and $v(p,q)$ in advance.

According to different assumptions, three models can be established: $\lambda_b=0$ and $\lambda_c\neq0$, referred to as the $cc$ model; $\lambda_b\neq0$ and $\lambda_c=0$, referred to as the $bc$ model; and $\lambda_b\neq0$ and $\lambda_c\neq0$, referred to as the $bcc$ model. 
These three models correspond to different physical pictures: the first retains only the direct interaction within the continuum, the second retains only the coupling between the bare state and the continuum, and the last retains both mechanisms simultaneously.

For the $bcc$ model, in addition to the phase shift, one also needs to know the energy at which $T(E_C)=0$, namely, the so-called Castillejo-Dalitz-Dyson zero. Clearly, we currently lack such data. 
On the other hand, according to Ref.~\cite{Li:2022aru}, for the $cc$ model the phase shift satisfies $\delta(0) - \delta(\infty) = -\pi$, which can also be observed from the data presented in Fig.~\ref{fig:phaseshiftdata}. Since there are no bound states, the Levinson theorem requires the existence of a bare state, which rules out the applicability of the $cc$ model. 
Therefore in the present work we focus on the $bc$ model.

\subsection{$bc$ model}

In the $bc$ model, only the coupling between the bare state $\rho_0$ and the $\pi\pi$ continuum is retained. 
The aim is to reconstruct the bare-state mass and the $\rho_0\pi\pi$ vertex strength from the $P$-wave $\pi\pi$ phase shift. 
Following Ref.~\cite{Li:2022aru}, they are given by
\begin{align}
m_0&=E_{\mathrm{th}}-\frac{1}{\pi}\int_{E_{\mathrm{th}}}^{\infty} dE\delta(E),\label{eq:traceformula}
\end{align}
\begin{align}
\lambda_b^2f^2(p)
&=\frac{(2\pi)^3h'_p |\sin\delta(h_p)|(h_p-E_{\mathrm{th}})}{\pi p^2}
\nonumber \\
&~~~~~~~~~~~\times \exp\bigg(\frac{1}{\pi}\mathcal{P}\int_{E_{\mathrm{th}}}^{\infty} dE \frac{\delta(E)}{h_p-E}\bigg)
.
\label{eq:interaction}
\end{align}
Here $h'_p=d h_p/dp$, and $\mathcal{P}\int dE$ denotes the principal-value integral. 
Eq.~\eqref{eq:interaction} shows that the vertex strength $\lambda_b^2 f^2(p)$ is fixed by the phase shift through both the factor $|\sin\delta(h_p)|$ and the principal-value integral.

\begin{figure*}[hpbt]
\centering
\includegraphics[width=0.85\textwidth]{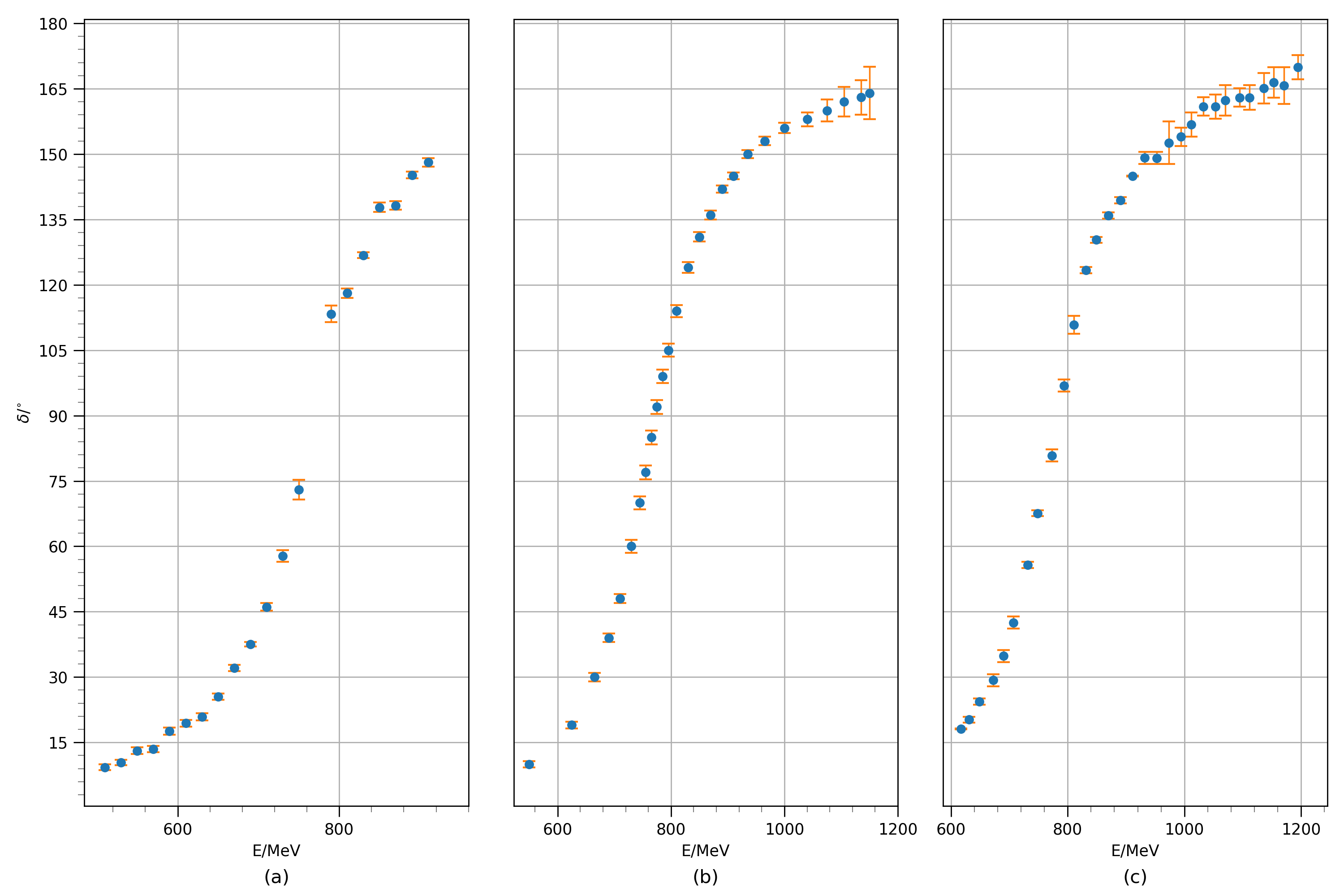}
\caption{$P$-wave $\pi\pi$ phase shifts in the $I=1$ channel. 
The data in panels (a), (b), and (c) are taken from Refs.~\cite{Estabrooks1974}, \cite{PhysRevD.7.1279}, and \cite{HYAMS1973134}, respectively.}
\label{fig:phaseshiftdata}
\end{figure*}

There are two practical limitations in applying these relations to the $\rho$ meson. 
First, the inverse-scattering reconstruction determines $\lambda_b^2 f^2(p)$ only on the real energy axis. 
Thus, Eq.~\eqref{eq:interaction} cannot be reliably extrapolated to the complex energy plane, and the resonance pole cannot be obtained directly in the present construction. 
Second, the experimental phase-shift data cover only a finite-energy interval, as shown in Fig.~\ref{fig:phaseshiftdata}. 
The data are limited near threshold and unavailable at sufficiently high energies. 
Moreover, when the energy becomes too high, additional reaction channels open, whereas the present derivation is based on a single-channel assumption.
For these reasons, we restrict the analysis to the region below $1.2~\mathrm{GeV}$.

The bare mass extracted from the quark model provides an additional constraint on the unknown high-energy part of the phase-shift integral. 
The upper limit of the integration is denoted by $\Lambda$. If the available experimental phase-shift data do not extend to an energy of $1.2~\mathrm{GeV}$, we set $\Lambda = E_{\text{max}}$, where $E_{\text{max}}$ represents the maximum accessible energy. Otherwise, we set $\Lambda = 1.2~\mathrm{GeV}$. This convention for $\Lambda$ is adopted uniformly hereafter. For $\Lambda \ge h_p$, one obtains
\begin{align}
&\mathcal{P}\int_{E_{\mathrm{th}}}^{\Lambda} dE \frac{\delta(E)}{E-h_p}
>\mathcal{P}\int_{E_{\mathrm{th}}}^{\infty} dE \frac{\delta(E)}{E-h_p}\notag\\
>&\mathcal{P}\int_{E_{\mathrm{th}}}^{\Lambda} dE \frac{\delta(E)}{E-h_p}
+\frac{\pi(E_{\mathrm{th}}-m_0)-\int_{E_{\mathrm{th}}}^{\Lambda} dE\delta(E)}{\Lambda - h_p}.
\label{eq:integrationuplow}
\end{align}
Here the sign convention for the phase shift should be specified. 
Following Ref.~\cite{Li:2022aru} and Levinson's theorem~\cite{PhysRevC.45.418}, we take $\delta(E_{\mathrm{th}})=-\pi$ for the $\rho$-meson system, where no bound state is present. 
With this convention, the phase shift entering the present single-channel analysis is negative in the energy region considered. 
Using Eq.~\eqref{eq:integrationuplow} in Eq.~\eqref{eq:interaction}, one obtains the upper and lower bounds of $\lambda_b^2 f^2(p)$. 
Thus, although the high-energy phase shift is not known, its effect on the reconstructed interaction can be restricted to a finite range by the bare-mass constraint.

For the present $bc$ model, the $T$-matrix expression of Ref.~\cite{Li:2022aru} gives
\begin{align}
\label{eq:Tmatrix}
&\bra{q}T(E+i\varepsilon)\ket{q}\notag\\
&=\frac{\lambda_b^2f^2(q)}
{E-m_0-\mathcal{P}\int\frac{k^2dk}{(2\pi)^3}
\frac{\lambda_b^2f^2(k)}{E-h_k}
+i\pi\frac{q^2 \lambda_b^2f^2(q)}{(2\pi)^3h_q'}} .
\end{align}
Equation~\eqref{eq:Tmatrix} differs from both the standard Breit-Wigner form and the Flatt\'e parametrization. 
The denominator contains the bare-mass term, the real self-energy generated by the $\pi\pi$ continuum, and the on-shell imaginary part fixed by the $\pi\pi$ phase space. 
Strictly speaking, the resonance parameters should be extracted from the pole position of the $T$-matrix on the proper unphysical sheet of the complex energy plane. 
In the present inverse-scattering construction, however, the interaction strength $\lambda_b^2 f^2(q)$ is reconstructed only on the real energy axis from the $P$-wave $\pi\pi$ phase shift, and its analytic continuation away from the real axis is not fixed by the inversion. 
Thus, the pole position cannot be determined directly within the present setup. 
We therefore use two approximate prescriptions to estimate the effective Breit-Wigner width of the $\rho$ meson.

The first approach is the most direct Breit-Wigner type approximation, namely, one regards the imaginary part in the denominator of Eq.~\eqref{eq:Tmatrix} as one half of the Breit-Wigner width, and the real part as the Breit-Wigner mass term. 
One then defines
\begin{align}
	&\Gamma_0^{\mathrm{BW}}(q) 
	=\frac{\pi q^2}{(2\pi)^2 h_q'}\lambda_b^2 f^2(q)\notag\\
	&=2\exp\left(\frac{1}{\pi}\mathcal{P}\int dE \frac{\delta(E)}{h_q-E}\right)
    |\sin\delta(h_q)|(h_q-E_{\mathrm{th}}).
	\label{eq:width}
\end{align}
With the phase-shift convention used in this work, the resonance point is determined by $\delta(M)=-\pi/2$. 
We then define the $\rho$-meson width in this prescription as the value of $\Gamma_0^{\mathrm{BW}}(q)$ at this point, where
\begin{equation}
	q=\sqrt{\frac{M^2}{4}-m_{\pi}^2}.
\end{equation}
Here $M$ denotes the energy at which the phase shift reaches $-\pi/2$. 
Combining Eqs.~\eqref{eq:integrationuplow} and~\eqref{eq:width}, the allowed range of $\Gamma$ can be obtained. 
This definition is the most direct one, but it does not explicitly account for the energy dependence of the principal-value integral.

Second, more rigorous consideration is based on the following derivation.
Consider the scattering amplitude $T(E)$ in the single-channel case and write its inverse as
\begin{equation}
	T^{-1}(E)=A(E)+iB(E),
	\label{eq:Tm1AB}
\end{equation}
where $A(E)$ and $B(E)$ are real functions on the real energy axis. 
The resonance mass parameter $M$ is defined by the zero of the real part of the inverse amplitude,
\begin{equation}
	A(M)=0 .
	\label{eq:am0}
\end{equation}
It should be emphasized that, with our phase-shift convention, the physical $P$-wave $\pi\pi$ phase shift carries a negative sign. 
Therefore, the condition $A(M)=0$ corresponds to $\delta(M)=-\pi/2$, rather than $+\pi/2$. 
Around $E=M$, the real part can be expanded to first order as
\begin{equation}
	A(E)\simeq A'(M)(E-M),
\end{equation}
where $A'(M)=dA/dE|_{E=M}$. 
For an isolated narrow resonance, the imaginary part varies slowly in the same energy region and can be approximated by its value at $E=M$,
\begin{equation}
	B(E)\simeq B(M).
\end{equation}
The amplitude then takes the Breit-Wigner-like form
\begin{align}
	T(E)
	&\simeq \frac{1}{A'(M)(E-M)+iB(M)} \notag\\
	&= \frac{1}{A'(M)}
	\frac{1}{(E-M)+i\,B(M)/A'(M)} .
\end{align}

The Breit-Wigner form used for this comparison is written as
\begin{equation}
	T_{\mathrm{BW}}(E)=\frac{g}{(E-M)+i\,\Gamma/2}.
	\label{eq:Tbw}
\end{equation}
Matching this expression to the inverse-amplitude expansion gives
\begin{equation}
	\frac{\Gamma}{2}=\frac{B(M)}{A'(M)}.
\end{equation}
For the $bc$ model, comparison between Eq.~\eqref{eq:Tmatrix} and Eq.~\eqref{eq:Tm1AB} gives
\begin{align}
	A(E)&=\frac{E-m_0-\mathcal{P}\int\frac{k^2dk}{(2\pi)^3}\frac{\lambda_b^2 f^2(k)}{E-h_k}}{\lambda_b^2 f^2(q)},
	\\
	B(E)&=\pi\frac{q^2}{(2\pi)^3h_q'}.
\end{align}
Thus,
\begin{align}
	A'(M)=
	\frac{
		1-\left.\frac{d}{dE}\left(\mathcal{P}\int \frac{k^2dk}{(2\pi)^3}
		\frac{\lambda_b^2 f^2(k)}{E-h_k}\right)
		\right|_{E=M}
	}{\lambda_b^2 f^2(q)}.
\end{align}
Here the resonance condition in Eq.~\eqref{eq:am0} has been used, which gives
\begin{align}
M-m_0-\mathcal{P}\!\int \frac{k^2dk}{(2\pi)^3}\frac{\lambda_b^2 f^2(k)}{M-h_k}=0.
\label{eq:halfpiphase}
\end{align}

The derivative of the principal-value term can be separated into the experimentally constrained part and the high-energy remainder,
\begin{align}
	&\frac{d}{dE}\left.\left(\mathcal{P}\int \frac{k^2dk}{(2\pi)^3}
	\frac{\lambda_b^2 f^2(k)}{E-h_k}\right)\right|_{E=M}\nonumber\\
    =&\frac{d}{dE}\left.\left(\mathcal{P}\int^{k_{\Lambda}}_{0} \frac{k^2dk}{(2\pi)^3}
	\frac{\lambda_b^2 f^2(k)}{E-h_k}\right)\right|_{E=M}
    \nonumber\\
    &~~~~~~~~~~~~~~~~~~~~~~~~
    +
    \int^\infty_{k_{\Lambda}}\frac{k^2dk}{(2\pi)^3}
	\frac{-\lambda_b^2 f^2(k)}{(M-h_k)^2}.
	\label{eq:ddE}
\end{align}
Here $h_{k_{\Lambda}}=\Lambda$, and $\Lambda$ denotes the highest energy covered by the experimental phase-shift data or $1.2\ \mathrm{GeV}$. 
Since $\lambda_b^2 f^2(k)$ in the interval $[0,k_{\Lambda}]$ is determined from Eq.~\eqref{eq:interaction}, the first term on the right-hand side of Eq.~\eqref{eq:ddE} can be evaluated numerically. 
The second term depends on the unknown high-energy part of the interaction. 
Its contribution can nevertheless be bounded as
\begin{align}
	0>&
    \int^\infty_{k_{\Lambda}}
	\frac{-\lambda_b^2 f^2(k)k^2dk}{(2\pi)^3(M-h_k)^2}
    >\frac{1}{\Lambda-M}\int^\infty_{k_{\Lambda}} \frac{k^2dk}{(2\pi)^3}
	\frac{\lambda_b^2 f^2(k)}{M-h_k}
    \nonumber\\
=&\frac{1}{\Lambda-M}\left(M-m_0-\mathcal{P}\int^{k_{\Lambda}}_{0} \frac{k^2dk}{(2\pi)^3}\frac{\lambda_b^2 f^2(k)}{M-h_k}\right).
\label{eq:ddEinfty}
\end{align}
In the last step, Eq.~\eqref{eq:halfpiphase} has been used to replace the high-energy integral by the resonance condition. 
Combining the above relations, the corrected Breit-Wigner width is obtained as
\begin{equation}
	\Gamma_1^{\mathrm{BW}}
	=
	\frac{\Gamma_0^{\mathrm{BW}}}{1-\frac{d}{dE}\left.\left(\mathcal{P}\int \frac{k^2dk}{(2\pi)^3}\frac{\lambda_b^2 f^2(k)}{E-h_k}\right)\right|_{E=M}}.
	\label{eq:Gamma1bw}
\end{equation}
The denominator in Eq.~\eqref{eq:Gamma1bw} represents the correction from the energy dependence of the principal-value self-energy. 
Equivalently, it plays the role of a loop-renormalization factor. 
In this sense, the coupling $g$ in Eq.~\eqref{eq:Tbw} is the dressed coupling appearing in the Breit-Wigner-like amplitude, while $\lambda_b^2 f^2(q)$ is the corresponding bare vertex strength in the $bc$ model.

Finally, the components of $\rho$ meson are also explored here.
Theoretically, the $\rho$ meson is usually regarded as a pole of the analytically continued $T$-matrix in the complex plane; experimentally, what can be observed is the ``projection'' of this pole onto the real axis, namely, the resonance peak characterized by the Breit-Wigner parametrization. 
From this perspective, the $\rho$ meson is essentially a broad resonance peak on the real axis, which is collectively composed of many scattering states $\ket{p^+}$ on the real axis.
The Breit-Wigner curve may be understood as the weight distribution of these scattering states at different energy points. 
Therefore, we describe the projection of the physical $\rho$ on the real axis by
\begin{equation}
\ket{\rho}=\int dE\sqrt{\frac{2\Gamma}{\pi}}\frac{E}{|E^2-m_{\rho}^2+im_{\rho}\Gamma|}\ket{\psi^+_{\pi\pi}(E)},
\label{eq:rhophysical}
\end{equation}
where $\ket{\psi_{\pi\pi}^+(E)} = \frac{\sqrt{E^2-4m_{\pi}^2}}{2(2\pi)^{3/2}} \left(\frac{E}{2\sqrt{E^2-4m_{\pi}^2}}\right)^{1/2}\ket{p^+}$.
The definition in Eq.~\eqref{eq:rhophysical} satisfies $\mathrm{tr}\,\hat{\rho}=1$ through $\hat{\rho}\equiv \ket{\rho}\bra{\rho}$. According to
\begin{equation}
	\ket{p^+}=\ket{p}+\frac{1}{h_p-H_0+i\varepsilon}H_I\ket{p^+},
\end{equation}
the coupling between the bare state $\ket{\rho_0}$ and $\ket{p^+}$ can be calculated by the following expression (for the detailed derivation, see Ref.~\cite{Li:2022aru}):
\begin{align}
&|\langle \rho_0|p^+\rangle|^2 \nonumber \\
&= \frac{(2\pi)^3 h'_p}{\pi p^2} \exp \left( \frac{1}{\pi} \mathcal{P} \int dE \frac{\delta(E)}{E-h_p} \right) \frac{|\sin \delta(h_p)|}{h_p - E_{\mathrm{th}}}.
\label{eq:bare_portion}
\end{align}
Combined with Eq.~\eqref{eq:rhophysical}, one can estimate the fraction of the bare $\rho_0$ component in the $\rho$, namely, $|\langle \rho|\rho_0\rangle|^2$. 
This reflects the fraction of the pure quark-gluon component in the physical $\rho$ meson.

\section{Results and discussion}
\label{sec:results}

We now extract the physical information of the $\rho$ meson according to the formulas given in the previous section.
First, we collect the experimental $P$-wave phase-shift data of $\pi\pi$ scattering, as shown in Fig.~\ref{fig:phaseshiftdata}, with the data taken from Refs.~\cite{Estabrooks1974,PhysRevD.7.1279,HYAMS1973134}. 
Among the different datasets, the lowest energies in Refs.~\cite{HYAMS1973134}, \cite{PhysRevD.7.1279}, and \cite{Estabrooks1974} are about $600$, $550$, and $500~\mathrm{MeV}$, respectively, all of which are far above the $\pi\pi$ threshold (about $280~\mathrm{MeV}$). 
Since the phase shift at threshold is known to be $-\pi$, in this work we directly connect the threshold point and the lowest experimental point with a straight line, and then treat the three datasets independently.
A continuous phase-shift curve is then obtained via linear interpolation over the interval $[E_{\mathrm{th}},\ 1.2\ \mathrm{GeV}]$, or up to the maximum available energy $\Lambda$ if the data do not reach $1.2\,\mathrm{GeV}$. 
Compared with the interval $[1.2~\mathrm{GeV},\ \infty)$, the unknown part in the former interval is confined only to a finite small region near threshold, and therefore the extrapolation based on the existing data is relatively reliable and will not introduce significant errors; by contrast, for the latter interval, the unknown part extends infinitely and cannot be controlled by the available data. 
Overall, within a controllable error range, we may regard the phase shift in the range $[E_{\mathrm{th}},\ 1.2\ \mathrm{GeV}]$ as known.

While the overall interpolation error is controllable, the linear connection over the energy axis introduces a specific deviation in the extreme threshold limit. 
For $P$-wave $\pi\pi$ scattering, the near-threshold phase shift should strictly scale as $p^3$. 
In contrast, the linear energy interpolation effectively reduces this scaling to $p^2$. 
Nevertheless, because our primary focus is on the $\rho$ meson resonance region, and we aim to extract the interaction directly from experimental data without introducing additional background models, this data-driven procedure remains a practical and objective choice. 
We expect that future precision measurements near the threshold will naturally refine the theoretical description of this interaction.
For these three sets of phase-shift data, we carry out the calculations independently.

\begin{figure*}[t]
\centering
\includegraphics[width=\textwidth]{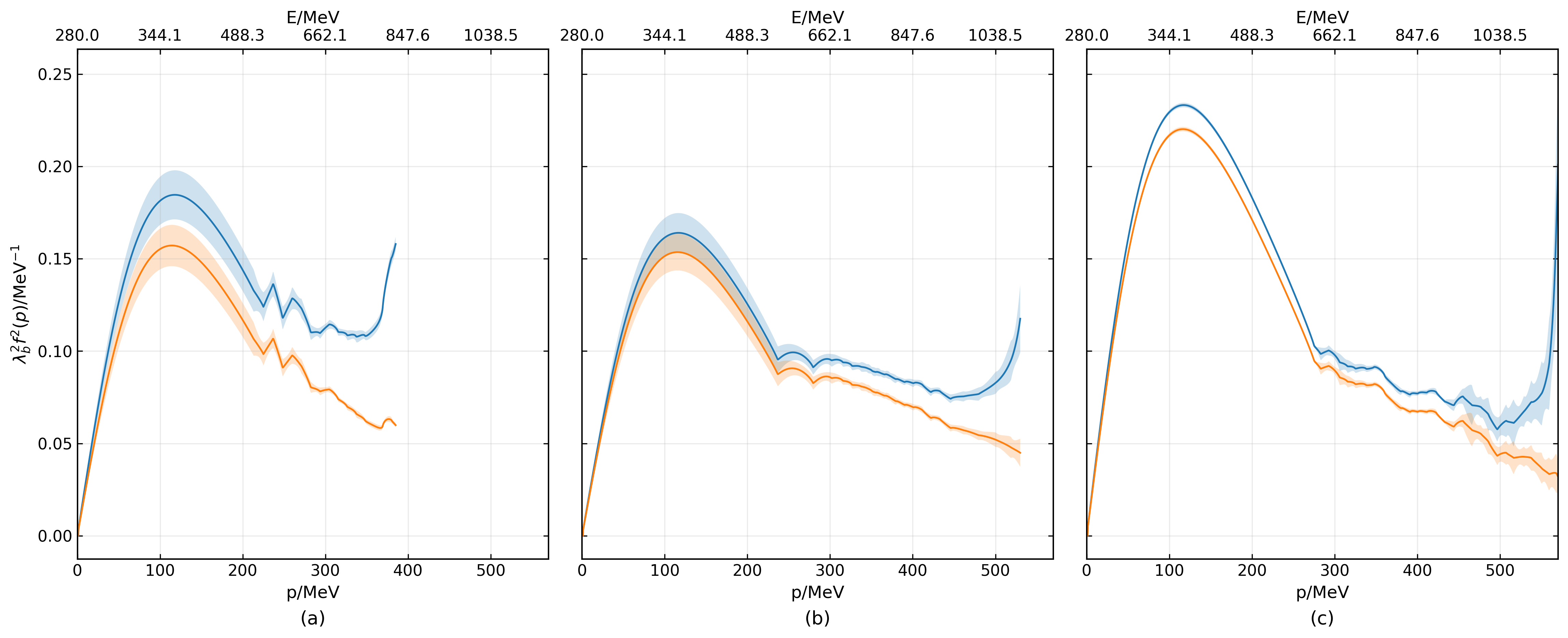}
\caption{
The $\rho_0\pi\pi$ interaction strength $\lambda_b^2 f^2(p)$. 
Panels (a), (b), and (c) are obtained from the phase-shift data shown in Figs.~\ref{fig:phaseshiftdata}(a), \ref{fig:phaseshiftdata}(b), and \ref{fig:phaseshiftdata}(c), respectively.
The lower horizontal axis denotes the pion momentum $p$ in the center-of-mass frame, while the upper axis gives the corresponding total $\pi\pi$ energy, $E=h_p=2\sqrt{m_\pi^2+p^2}$. 
The blue and orange curves denote the lower and upper bounds, respectively, and the shaded bands represent the corresponding uncertainties. 
The small irregularities of the curves arise from the linear interpolation of the discrete phase-shift data.
}
\label{fig:lambda2f2}
\end{figure*}

Based on the above phase-shift curves and Eqs.~\eqref{eq:interaction} and \eqref{eq:integrationuplow}, we plot the upper and lower bounds of $\lambda_b^2 f^2(p)$ in \figref{fig:lambda2f2}, where the error bands are evaluated using the bootstrap method.

The results in the figure show that, under the existing phase-shift data and the bare-mass constraint, the coupling strength between $\rho_0$ and the $\pi\pi$ channel can be restricted to a finite interval.
It should be pointed out that when the energy approaches the upper limit $\Lambda$, the upper bound of $\lambda_b^2 f^2(p)$ diverges to infinity due to the factor $1/(\Lambda-h_p)$ in the expression.

For the width, in the $bc$ model, let us first estimate the difference between $\Gamma^{\text{BW}}_0$ and $\Gamma^{\text{BW}}_1$.
We can estimate the values of 
\begin{align}
\frac{d}{dE}\left.\left(\mathcal{P}\int \frac{k^2dk}{(2\pi)^3}\frac{\lambda_b^2 f^2(k)}{E-h_k}\right)\right|_{E=M},
\end{align}
for the three datasets as follows: approximately $-0.62 \sim -0.55$ for \figref{fig:phaseshiftdata}(a), $0.032 \sim 0.045$ for \figref{fig:phaseshiftdata}(b), and $-0.029 \sim -0.028$ for \figref{fig:phaseshiftdata}(c).
Notably, the result calculated from the data in Fig.~\ref{fig:phaseshiftdata}(a) differs drastically from the latter two sets.
The primary reason for this discrepancy is that the upper energy limit of this dataset is exceedingly low, leading to a disproportionately large impact from the factor $\frac{1}{\Lambda-E}$ in Eq.~\eqref{eq:ddEinfty}.
Obviously, this value is rather small compare to 1 in Eq.~(\ref{eq:Gamma1bw}).
So we take the approximation as $\Gamma^{\text{BW}}_0=\Gamma^{\text{BW}}_1$ in the $bc$ model.
We show the results in Table~\ref{tab:results}.

Subsequently, using Eqs.~(\ref{eq:rhophysical})--(\ref{eq:bare_portion}), we calculate the ranges of the overlap between the bare state and the physical state for various cases, as shown in Table~\ref{tab:results}.
It can be seen from Table~\ref{tab:results} that the results obtained from different experimental datasets are overall compatible with each other, indicating that the method exhibits a certain degree of stability against variations in the input data. 
On the other hand, there remain noticeable differences between the upper- and lower-bound solutions; however, this discrepancy is expected to be further reduced as more complete phase-shift information becomes available. Under the current data and model assumptions, a reasonable interval estimate for the width of the $\rho$ meson and the bare-state component can be obtained, although the result is not highly precise.

Furthermore, under our phase-shift convention where $\delta(E)<0$, in combination with Eq.~\eqref{eq:traceformula}, we obtain
\begin{equation*}
    m_0 > E_{\mathrm{th}} - \frac{1}{\pi} \int_{E_{\mathrm{th}}}^{\Lambda} dE\, \delta(E).
\end{equation*}
In view of the experimental data employed in this work, we present the lower bound for the bare mass: $m_0 > 757.6~\mathrm{MeV}$ for the result in \figref{fig:phaseshiftdata}(a); $m_0 > 793.6~\mathrm{MeV}$ for the result in \figref{fig:phaseshiftdata}(b); and $m_0 > 797.5~\mathrm{MeV}$ for the result in \figref{fig:phaseshiftdata}(c). 
Since the third set of results provides the most extensive data range, reaching up to $1.2~\mathrm{GeV}$, it yields the most stringent constraint. 
Based on this constraint, the bare mass should be greater than $800~\mathrm{MeV}$, which is largely consistent with the findings in Ref.~\cite{Li:2022aru} and aligns with the expectations derived from the quark model.
Correspondingly, it can be seen from Eq.~(\ref{eq:integrationuplow}) that, when the bare mass reaches its minimum value, the upper and lower limits of the width coincide.
To a certain extent, this corroborates the interpretation adopted in the present work, namely, that the eigenvalues of the quark model correspond to the bare mass.

\begin{table}[htbp]
\caption{The decay widths $\Gamma$ (in MeV) derived from three sets of phase-shift data (see \figref{fig:phaseshiftdata}), as well as the bare-mass contribution to the $\rho$-meson system, are calculated using the $bc$ model framework. The corresponding statistical uncertainties are provided for each individual result.
}
\label{tab:results}
\begin{ruledtabular}
\begin{tabular}{@{}cccc@{}}
Dataset & Solution & $\Gamma^{\text{BW}}_0 \pm \Delta\Gamma^{\text{BW}}_0$ (MeV) & $|\langle \rho|\rho_0\rangle|^2 \pm \Delta |\langle \rho|\rho_0\rangle|^2$ \\
\colrule
\multirow{2}{*}{\figref{fig:phaseshiftdata}(a)}
& Lower & $104.0 \pm 2.5$ & $0.31 \pm 0.0036$ \\
& Upper & $192.0 \pm 4.6$ & $0.82 \pm 0.0028$ \\
\colrule
\multirow{2}{*}{\figref{fig:phaseshiftdata}(b)}
& Lower & $135.1 \pm 2.8$ & $0.72 \pm 0.0042$ \\
& Upper & $154.9 \pm 3.4$ & $0.88 \pm 0.0023$ \\
\colrule
\multirow{2}{*}{\figref{fig:phaseshiftdata}(c)}
& Lower & $135.3 \pm 2.7$ & $0.75 \pm 0.0030$ \\
& Upper & $151.9 \pm 3.0$ & $0.89 \pm 0.0018$
\end{tabular}
\end{ruledtabular}
\end{table}

\section{Summary and outlook}
\label{sec:sum}

This work is based on the following viewpoint: since the interaction in the quark model acts mainly at the quark-gluon level, the hadron mass calculated from this model may be regarded as the corresponding bare mass.
On this basis, we take the bare mass provided by the quark model as input and, together with the experimentally measured scattering phase-shift data, construct the corresponding model within the inverse-scattering theory framework, thereby extracting the physical width, bare-state component, and other related physical information of the $\rho$ meson.

In the concrete calculations, we first refit the parameters of the quark model and improve the form of $\alpha_s(Q^2)$ in different energy regions so as to describe more reasonably the effective interaction in the constituent quark system.
Subsequently, using the new fitting parameters, we obtain the bare mass of the $\rho$ meson and complete the numerical calculation of the meson spectrum with GEM.
On this basis, we further construct the $bc$ model in inverse-scattering theory and discuss in detail the relation among the scattering phase shift, the bare mass, and the physical properties of the $\rho$ meson.

By analyzing the experimental $P$-wave $\pi\pi$ scattering phase-shift data, we obtain the relevant coupling quantity of the $\rho$ meson, the range of its physical width, and the overlap information between the bare state and the physical state.
The results show that combining the bare-state input at the quark level with the scattering information at the hadronic level can provide a new way of describing the structure of the $\rho$ meson and its resonant properties. 
This also indicates that the bare-mass information provided by the quark model still has important reference value for the study of actually observed hadronic resonances.

Looking forward, this theoretical framework can still be further extended and improved. 
For example, it may be generalized to more hadronic systems with significant channel-coupling effects, and more complete dynamical information such as multichannel coupling and pole structures in the complex energy plane may also be incorporated, so as to reveal more deeply the internal structure and formation mechanism of hadronic resonances.

\section*{ACKNOWLEDGMENTS}
This work is supported by the National Natural Science Foundation of China (NSFC) under Grant No. 12221005 and by the Chinese Academy of Sciences under Grant No. YSBR-101.

\section*{DATA AVAILABILITY}
The data that support the findings of this article are not publicly available. The data are available from the authors upon reasonable request.

\appendix
\section{Mixing Angle}\label{MixingAngle}
If the flavor eigenstates are expressed in terms of $\psi_1$ and $\psi_8$, and the mass eigenstates are represented by $\eta$ and $\eta'$, then
\begin{equation}
\psi_8=\frac{1}{\sqrt{6}}\bigl(u\bar u + d\bar d - 2 s\bar s\bigr),
\end{equation}
\begin{equation}
\psi_1=\frac{1}{\sqrt{3}}\bigl(u\bar u + d\bar d + s\bar s\bigr),
\end{equation}
\begin{equation}
\eta=\cos\theta\ \psi_8-\sin\theta\ \psi_1,
\end{equation}
\begin{equation}
\eta'=\sin\theta\ \psi_8+\cos\theta\ \psi_1.
\end{equation}

In the case of ideal mixing, one can introduce $\psi_n$ and $\psi_s$,
\begin{equation}
\psi_n=\frac{1}{\sqrt{2}}(u\bar u+d\bar d),\quad\psi_s=s\bar s,
\end{equation}
leading to the following expressions:
\begin{equation}
\psi_8=\cos\alpha\ \psi_n-\sin\alpha\ \psi_s,
\end{equation}
\begin{equation}
\psi_1=\sin\alpha\ \psi_n+\cos\alpha\ \psi_s,
\end{equation}
where $\alpha \approx 54.7^\circ$.

Then the mass eigenstates can be expressed as
\begin{equation}
\eta = \cos(\alpha+\theta)\  \psi_n -\sin(\alpha+\theta)\  \psi_s,
\label{realmixformula1}
\end{equation}
\begin{equation}
\eta' = \sin(\alpha+\theta)\  \psi_n + \cos(\alpha+\theta)\ \psi_s.
\label{realmixformula2}
\end{equation}
In the above expressions, \(\theta\) is the desired mixing angle. This approach is used because it avoids the problem of \(\psi_8\) and \(\psi_1\) not being mass eigenstates in the Hamiltonian framework, making it impossible to determine their eigenvalues directly. According to Eqs.~\eqref{realmixformula1} and \eqref{realmixformula2}, solving for the eigenvalues of \(\psi_8\) and \(\psi_1\) is transformed into solving for the eigenvalues of \(\psi_n\) and \(\psi_s\). The mixing angle can then be derived from the matrix elements of a $2\times2$ matrix.

\section{Stability of the bare-mass inputs in the quark model fit}
\label{app:fit-stability}

In the main fit, $\rho$, $K^*$, and $D^*$ are not included in the mass input set. The eigenvalues obtained for these states from the quark model Hamiltonian are then read as bare masses. These bare masses should be distinguished from the measured masses, because the latter already contain shifts from open channels. To check the stability of this assignment, we include the three states again in the input set and repeat the global fit in two ways. The two refits show whether the bare masses obtained in the main fit are stable, and how strongly the spectrum changes when the measured physical masses are used as mass inputs in the same bare Hamiltonian. For simplicity, the check is carried out with model~1. Since models~1--3 have the same potential form and differ only in the parametrization of $\alpha_s$, the same qualitative behavior is expected for models~2 and 3.

Fit~1 uses the bare masses obtained in the main fit as the three input masses. Following the notation $\rho_0$ for the bare $\rho$ state, the three masses are denoted by
\begin{equation}
\begin{gathered}
M_{\rho_0}=846.8~\mathrm{MeV},\qquad
M_{(K^*)_0}=950.4~\mathrm{MeV},\\
M_{(D^*)_0}=2027.2~\mathrm{MeV}.
\end{gathered}
\end{equation}
Fit~2 uses the measured physical masses of the same states. The two fits use the same Hamiltonian and the same remaining input set. Only the three restored masses are changed, so the comparison shows how the spectrum responds to the bare-mass input and to the physical-mass input.

Tables~\ref{tab:app-fit-parameters} and \ref{tab:app-fit-spectrum} list the fitted parameters and spectra. In fit~1, the parameters and masses stay close to the main fit. With the same target error $\sigma=5~\mathrm{MeV}$, the mass deviation measure changes only from $\chi^2=16.9$ to $\chi^2\simeq17.2$. The added bare-mass inputs do not change the quark model determination.

Fit~2 gives a very different result. The mass deviation measure rises to $\chi^2\simeq141$, and the $\omega$ mass moves to $744.5~\mathrm{MeV}$, far below the measured value. Using the physical masses of $\rho$, $K^*$, and $D^*$ as inputs pulls the remaining narrow spectrum away from experiment. The comparison supports the treatment adopted in the main text. The quark model eigenvalues of these states should be regarded as bare masses. Their measured masses should be obtained only after the mass shifts from open channels are included.

\begin{table}[htpb]
\caption{
Quark model parameters from the two auxiliary fits. Fit~1 uses the bare-mass inputs $M_{\rho_0}$, $M_{(K^*)_0}$, and $M_{(D^*)_0}$ from the main fit. Fit~2 uses their measured physical masses.
}
\label{tab:app-fit-parameters}
\setlength{\tabcolsep}{6pt}
\begin{ruledtabular}
\begin{tabular}{lrr}
\textbf{Parameter} & \textbf{Fit~1} & \textbf{Fit~2} \\
\hline
\textbf{Quark masses} & & \\
$m_u = m_d$ (MeV) & 295 & 295 \\
$m_s$ (MeV) & 440 & 488.4 \\
$m_c$ (MeV) & 1670.3 & 1719.5 \\
$m_b$ (MeV) & 5035.1 & 5081.8 \\
\hline
\textbf{Confinement} & & \\
$a_c$ (MeV) & 326.3 & 318.4 \\
$\mu_c$ (fm$^{-1}$) & 1.257 & 1.594 \\
$\Delta$ (MeV) & 170.6 & 212.2 \\
$a_s$ & 0.7148 & 0.6727 \\
\hline
\textbf{Goldstone bosons} & & \\
$m_{\pi}$ (fm$^{-1}$) & \multicolumn{2}{c}{0.70 (fixed)} \\
$m_{\eta}$ (fm$^{-1}$) & \multicolumn{2}{c}{2.77 (fixed)} \\
$m_{K}$ (fm$^{-1}$) & \multicolumn{2}{c}{2.51 (fixed)} \\
$\Lambda_{\pi}=\Lambda_{\sigma}$ (fm$^{-1}$) & \multicolumn{2}{c}{4.20 (fixed)} \\
$\Lambda_{\eta}=\Lambda_{K}$ (fm$^{-1}$) & \multicolumn{2}{c}{5.20 (fixed)} \\
$m_{\sigma}$ (fm$^{-1}$) & 3.574 & 4.15 \\
$g^2_{ch}/(4\pi)$ & 0.44876 & 0.43297 \\
$\theta_P$ ($^\circ$) & -20.6 & -21 \\
\hline
\textbf{OGE} & & \\
$\hat{r}_0$ (MeV fm) & 35.678 & 27.312 \\
$\hat{r}_g$ (MeV fm) & 25.315 & 19.002 \\
$\alpha_0$ & 0.5633 & 0.5538 \\
$\Lambda_0$ (MeV) & 258.7 & 207.5 \\
$c_1$ (MeV) & 1230 & 1108 \\
$c_2$ (MeV) & 1409 & 1447 \\
$k$ & 0.4686 & 0.5123 \\
\end{tabular}
\end{ruledtabular}
\end{table}

\begin{table*}[htpb]
\caption{
Meson spectra from the two auxiliary fits, compared with experimental data. In fit~1, the input masses for $\rho$, $K^*$, and $D^*$ are $M_{\rho_0}$, $M_{(K^*)_0}$, and $M_{(D^*)_0}$ from the main fit. The experimental masses of these three states are listed only for comparison.
}
\label{tab:app-fit-spectrum}
\setlength{\tabcolsep}{4pt}
\begin{ruledtabular}
\begin{tabular}{cccccc}
$n\,{}^{2S+1}L_J$ & $J^{P(C)}$ & State & Fit~1 & Fit~2 & Expt. \\
\colrule
$1\,{}^1S_0$ & $0^{-+}$ & $\pi$ & 135.9 & 136.2 & 135 \\
$1\,{}^1S_0$ & $0^{-+}$ & $\eta$ & 547.5 & 544.4 & 548 \\
$1\,{}^3S_1$ & $1^{--}$ & $\rho$ & 847.5 & 797.4 & 775 \\
$1\,{}^3S_1$ & $1^{--}$ & $\omega$ & 780.3 & 744.5 & 783 \\
$1\,{}^1S_0$ & $0^{-+}$ & $\eta'$ & 957.8 & 946.4 & 958 \\
$1\,{}^3S_1$ & $1^{--}$ & $\phi$ & 1023.0 & 1018.1 & 1019 \\
$1\,{}^1S_0$ & $0^{-}$ & $K$ & 496.7 & 496.5 & 498 \\
$1\,{}^3S_1$ & $1^{-}$ & $K^*$ & 950.9 & 922.3 & 896 \\
$1\,{}^1S_0$ & $0^{-}$ & $D$ & 1863.4 & 1863.8 & 1865 \\
$1\,{}^3S_1$ & $1^{-}$ & $D^*$ & 2027.5 & 2016.2 & 2007 \\
$1\,{}^1S_0$ & $0^{-}$ & $D_s$ & 1971.3 & 1971.6 & 1968 \\
$1\,{}^3S_1$ & $1^{-}$ & $D_s^*$ & 2112.5 & 2110.5 & 2112 \\
$1\,{}^1S_0$ & $0^{-+}$ & $\eta_c$ & 2983.8 & 2986.2 & 2984 \\
$1\,{}^3S_1$ & $1^{--}$ & $J/\psi$ & 3096.8 & 3098.0 & 3097 \\
$1\,{}^3P_0$ & $0^{++}$ & $\chi_{c0}$ & 3414.8 & 3414.8 & 3415 \\
$1\,{}^3P_1$ & $1^{++}$ & $\chi_{c1}$ & 3509.6 & 3510.2 & 3511 \\
$1\,{}^1P_1$ & $1^{+-}$ & $h_c$ & 3528.8 & 3530.6 & 3525 \\
$1\,{}^3P_2$ & $2^{++}$ & $\chi_{c2}$ & 3553.9 & 3554.6 & 3556 \\
$2\,{}^3S_1$ & $1^{--}$ & $\psi(2S)$ & 3685.0 & 3674.8 & 3686 \\
$1\,{}^1S_0$ & $0^{-+}$ & $\eta_b$ & 9395.0 & 9399.3 & 9399 \\
$1\,{}^3S_1$ & $1^{--}$ & $\Upsilon$ & 9465.3 & 9462.2 & 9460 \\
$1\,{}^3P_0$ & $0^{++}$ & $\chi_{b0}$ & 9858.6 & 9858.6 & 9859 \\
$1\,{}^3P_1$ & $1^{++}$ & $\chi_{b1}$ & 9890.3 & 9887.8 & 9893 \\
$1\,{}^1P_1$ & $1^{+-}$ & $h_b$ & 9898.5 & 9896.2 & 9899 \\
$1\,{}^3P_2$ & $2^{++}$ & $\chi_{b2}$ & 9909.4 & 9906.8 & 9912 \\
$2\,{}^3S_1$ & $1^{--}$ & $\Upsilon(2S)$ & 10035.3 & 10039.2 & 10023 \\
$1\,{}^1S_0$ & $0^{-}$ & $B$ & 5275.9 & 5281.7 & 5280 \\
$1\,{}^3S_1$ & $1^{-}$ & $B^*$ & 5333.7 & 5332.1 & 5325 \\
$1\,{}^1S_0$ & $0^{-}$ & $B_s$ & 5362.7 & 5363.1 & 5367 \\
$1\,{}^3S_1$ & $1^{-}$ & $B_s^*$ & 5413.9 & 5411.4 & 5415 \\
$1\,{}^1S_0$ & $0^{-}$ & $B_c$ & 6274.9 & 6274.3 & 6274 \\
$1\,{}^3D_2$ & $2^{--}$ & $\Upsilon_2$ & 10157.8 & 10159.3 & 10164 \\
\end{tabular}
\end{ruledtabular}
\end{table*}

\bibliography{refs_en}
\end{document}